# A charge-dependent electron emission yield from electron irradiated dielectrics

M. Belhaj[1], A.M. Cavalca de Almeida[1,2], N. Fil[2]

[1] DPHY, ONERA, Université de Toulouse, 31000, Toulouse, France

[2]CNES, 18 av. Edouard Belin, 31401 Toulouse CEDEX, France

Mohamed.belhaj@onera.fr

## Abstract

The total electron emission yield (TEEY) of dielectric surfaces may vary during electron irradiation owing to charge accumulation within the material. In the present work, the generation, transport, and escape formulation of Dionne is extended by introducing an additional attenuation of secondary electrons that depends on trapped positive charge. The resulting emission law is coupled to a charge-balance equation, allowing the TEEY to be calculated as a function of incident energy and injected charge. The model is examined using measurements on polycrystalline $Al_2O_3$, from which effective cross sections describing interactions between trapped charge and secondary electrons are determined. Their mean value is then used to predict the evolution of the TEEY curves without further adjustment. Finally, an analytical approximation of the nonlinear charging law is derived to facilitate the implementation in electron-irradiation simulations.

## 1. Introduction

Secondary-electron emission plays an important role in the interaction of energetic electrons with solid surfaces. It influences electron multipliers and vacuum detectors [1], spacecraft-material charging [2], and the onset of multipactor discharges in radio-frequency systems operating under vacuum [3]. The total electron emission yield (TEEY), denoted by σ, is defined as the number of emitted electrons per incident electron. It comprises the yield δ of secondary electrons generated within the solid and the yield η of backscattered primary electrons. Therefore, $\sigma(E_0) = \delta(E_0) + \eta(E_0)$, where $E_0$ is the primary-electron energy.

For insulating materials, the TEEY may evolve during irradiation because electron impacts and the associated emission processes lead to the accumulation of electrons and holes in the solid. The resulting space charge modifies the electric field both inside the material and in the adjacent vacuum. The external field affects electron trajectories and landing energies [4, 5, 6], whereas internal charging may alter secondary-electron generation, transport, and escape [7, 8, 9].

Although vacuum electrostatic fields can be included in particle simulations, the corresponding modification of emission processes within the dielectric is less commonly represented. The present work therefore addresses internal charging effects and their incorporation into calculations of electron exchange between irradiated dielectrics and their surroundings.

Several models describe electron emission for uncharged materials. Among them, the formulation of Dionne [10, 11] is particularly appropriate here because it separates secondary-electron generation from attenuation during transport and escape. Hence, the energy dependence of the yield can be related to the primary-electron penetration range and an effective attenuation length. In the original formulation, however, the transport parameters are assumed to remain unchanged during irradiation. Consequently, the model cannot describe charge-induced changes in emission without modification.

The effects of irradiation-induced charging on secondary-electron escape have been investigated experimentally and theoretically [12, 7]. Ganachaud and Mokrani [13] examined transport in internal electric fields, whereas Fitting and co-workers [14, 8] studied the attenuation and escape of low-energy electrons in insulators. These investigations indicate that charge-dependent emission may arise both from interactions with trapped charges and from changes in electron trajectories within the material. Accordingly, when a microscopic calculation is not required, the combined effects may be represented by an effective transport parameter.

More detailed descriptions solve the coupled problems of electron transport, trapping, recombination, and electrostatic-field evolution. Meyza and co-workers [15] developed a self-consistent model of charge transport and storage in alumina, while Touzin and co-workers [16] introduced a flight–drift model for electron-irradiated insulators. Such approaches provide spatial and temporal charge distributions. However, their computational requirements are substantially greater than those of a reduced emission law. A simpler description is therefore useful when the yield must be evaluated repeatedly in particle simulations, provided that its assumptions and effective parameters are specified.

Changes in electron emission arising from internal charging have been measured in several insulating solids. Measurements on polycrystalline MgO [9] and $Al_2O_3$ [17] showed a pronounced decrease in TEEY with increasing irradiation fluence under positive charging conditions. Positive charging in amorphous $SiO_2$ was also investigated experimentally and by Monte Carlo simulations [18]. Effective cross sections reported by Said et al. [19] for sapphire differ by approximately two orders of magnitude between positive ($\sim 10^{-9}$ cm$^2$) and negative ($\sim 10^{-11}$ cm$^2$) charging regimes. An even larger difference, of approximately four orders of magnitude, was reported for spinel by Boughariou et al. [20], with corresponding values of $\sim 10^{-10}$ and $\sim 10^{-14}$ cm$^2$. These effective parameters characterize different charge-related processes rather than directly comparable microscopic scattering mechanisms. Together with the reduced rate of yield evolution near the crossover energies [6], these observations motivate the present focus on the internal effects of excess trapped positive charge. The influence of internal electric fields on secondary-electron escape has also been examined analytically [21].

The distinction between external and internal charging becomes important when repeated electron impacts change the charge state of a dielectric surface. Particle simulations of spacecraft charging, such as SPIS, may include the electric field produced by accumulated charge and its effect on electron trajectories in vacuum [22]. However, this treatment does not necessarily account for the accompanying change in secondary-electron transport within the material.

Multipactor thresholds are sensitive to the TEEY law employed in the calculation [23]. Moreover, multipactor in partially dielectric-loaded microwave structures has been investigated both numerically and experimentally [24, 25]. A charge-dependent emission law is therefore required to represent internal dielectric-charging effects in such calculations without performing a microscopic transport simulation after each electron impact.

In the present model, the Dionne emission law is extended through an additional attenuation term associated with trapped positive charge. The resulting TEEY is coupled to a reduced positive-charge balance. This coupling gives a nonlinear feedback equation. An analytical approximation is therefore constructed. The predictions are then examined for polycrystalline $Al_2O_3$ and compared with those of the full nonlinear model.

## 2. Original Dionne model and proposed extension

### 2.1. Secondary-electron generation, transport and escape

In the treatment of Dionne, secondary-electron generation, transport through the solid, and escape into vacuum are considered as successive processes [10, 11]. For a one-dimensional geometry, the secondary-electron yield is

$$\delta_0(E_0) = B\int_0^{R(E_0)} g\,(x, E_0) exp(\frac{-x}{\lambda_0}) dx \quad (1)$$

Here x is the depth below the surface, $R(E_0)$ is the primary-electron penetration range, and $g(x, E_0)$ is the number of secondary electrons generated per unit depth and incident electron. The coefficient B represents effective escape through the surface, and $\lambda_0$ is the attenuation length in the reference material. The exponential term describes attenuation of secondary electrons generated at depth x due to elastic and inelastic interactions. The generation term is related to local energy deposition by

$$g(x, E_0) = \frac{1}{\varepsilon_{pair}} \frac{dE_{dep}(x, E_0)}{dx} \quad (2)$$

where $\varepsilon_{pair}$ is the mean energy required to create an electron–hole pair. If the energy deposited by the primary electron is taken as a first approximation to be distributed uniformly over its range, integration gives

$$\delta_0(E_0) = \frac{BE_0}{\varepsilon_{pair}} \frac{\lambda_0}{R(E_0)} [1 - exp(\frac{-R(E_0)}{\lambda_0})] \quad (3)$$

The range $R(E_0)$ may be calculated using an appropriate electron-penetration relation. In the present calculations for $Al_2O_3$, the Castaing expression [26] is employed:

$$R(E_0)(\mu m) = 0.033\frac{A/Z}{\rho}[E_0(keV)]^{1.7} \qquad (4)$$

In Eq. (4), A and Z are the atomic mass and atomic number, respectively, ρ is the mass density, and $E_0$ is expressed in keV. It is important to note that the Castaing expression was established for energies of the order of keV. Its use here at substantially lower energies (teens to hundreds eV) is an extrapolation adopted for the present calculations, rather than a range relation established in that energy interval.

The mean energy for electron–hole pair creation is approximately related to the band gap $E_g$ by the Alig–Bloom relation [27]:

$$\varepsilon_{pair} \approx 3E_g \qquad (5)$$

## 2.2. Effective attenuation associated with trapped charge

To account for changes in secondary-electron transport under irradiation, the reference attenuation length $\lambda_0$ is replaced by an effective attenuation length $\lambda_{eff}(D)$, dependent on the incident fluence D. The corresponding TEEY is obtained from the Dionne relation:

$$TEEY(E_0, D) = b + \frac{BE_0}{\varepsilon_{pair}}\frac{\lambda_{eff}(D)}{R(E_0)}[1 - exp(\frac{-R(E_0)}{\lambda_{eff}(D)})] \qquad (6)$$

The contribution of backscattered primary electrons is represented by b.

The effective interaction cross section $\Sigma_{eff}$ and the density $n_h$ of trapped holes determine the additional contribution to the inverse attenuation length:

$$\frac{1}{\lambda_{eff}} = \frac{1}{\lambda_0} + \Sigma_{eff}n_h \qquad (7)$$

The probability of a secondary electron generated at the depth x to reach the surface is then:

$$P(x, D) = \frac{1}{2}exp[-\int_0^x (\frac{1}{\lambda_0} + \Sigma_{eff}n_h(s, D))\, ds] \qquad (8)$$

s is a dummy variable of integration over depth.

## 2.3. Positive-charge balance and feedback

The emission yield determines the rate at which positive charge is trapped in the material. If $Q_t$ denotes the trapped charge and S the irradiated area, conservation of charge gives

$$\frac{d(Q_t/S)}{dD} = e[TEEY(E_0, D) - 1 - L(E_0, D)] \qquad (9)$$

The loss term L in Eq. (9) involves processes with distinct characteristic time scales. The first is rapid recombination of electron–hole pairs generated by ionization. Gibaru et al. [18] investigated the dependence of charge-carrier recombination on incident current density and the spatial overlap of electron cascades. At the low incident current densities considered here (less than 1 μA $cm^{-2}$), overlap between successive transient cascades is assumed to be negligible.

The second process concerns the relaxation of trapped positive charge. Experimental investigations have reported long-lived trapped-charge populations in alumina [28, 29], PTFE [30], amorphous $SiO_2$ [31], Kapton polyimide [33] (hundreds of seconds to hundreds of days). In the present model, the characteristic time of detrapping is assumed to be much longer than the characteristic irradiation time. Relaxation of trapped positive charge is therefore neglected, and no separate dynamic loss term is introduced in the charge-balance equation.

This assumption concerning detrapping and relaxation times is primarily applicable to bulk dielectric materials. In nanometre-scale coatings, other processes may substantially accelerate charge relaxation. The present model should therefore be applied with caution to thin films. Increasing temperature also accelerates detrapping.

Following the qualitative double-layer picture of electron-irradiated insulators proposed by Cazaux [31] in particular, the trapped-hole density $n_h$ is represented here by an effective population distributed within a near-surface layer of thickness $\lambda_0$. The low mean depth escape probability of secondary electrons motivates this representation [14, 32].

$$n_h = \frac{Q_t}{eS\lambda_0} \qquad (10)$$

where e is the elementary charge.

Equations (6)–(10) establish the feedback between secondary-electron emission and trapped charge. Positive charge increases secondary-electron attenuation, lowering the TEEY and thus reducing the rate of subsequent positive-charge accumulation. The feedback loop is summarized schematically below.

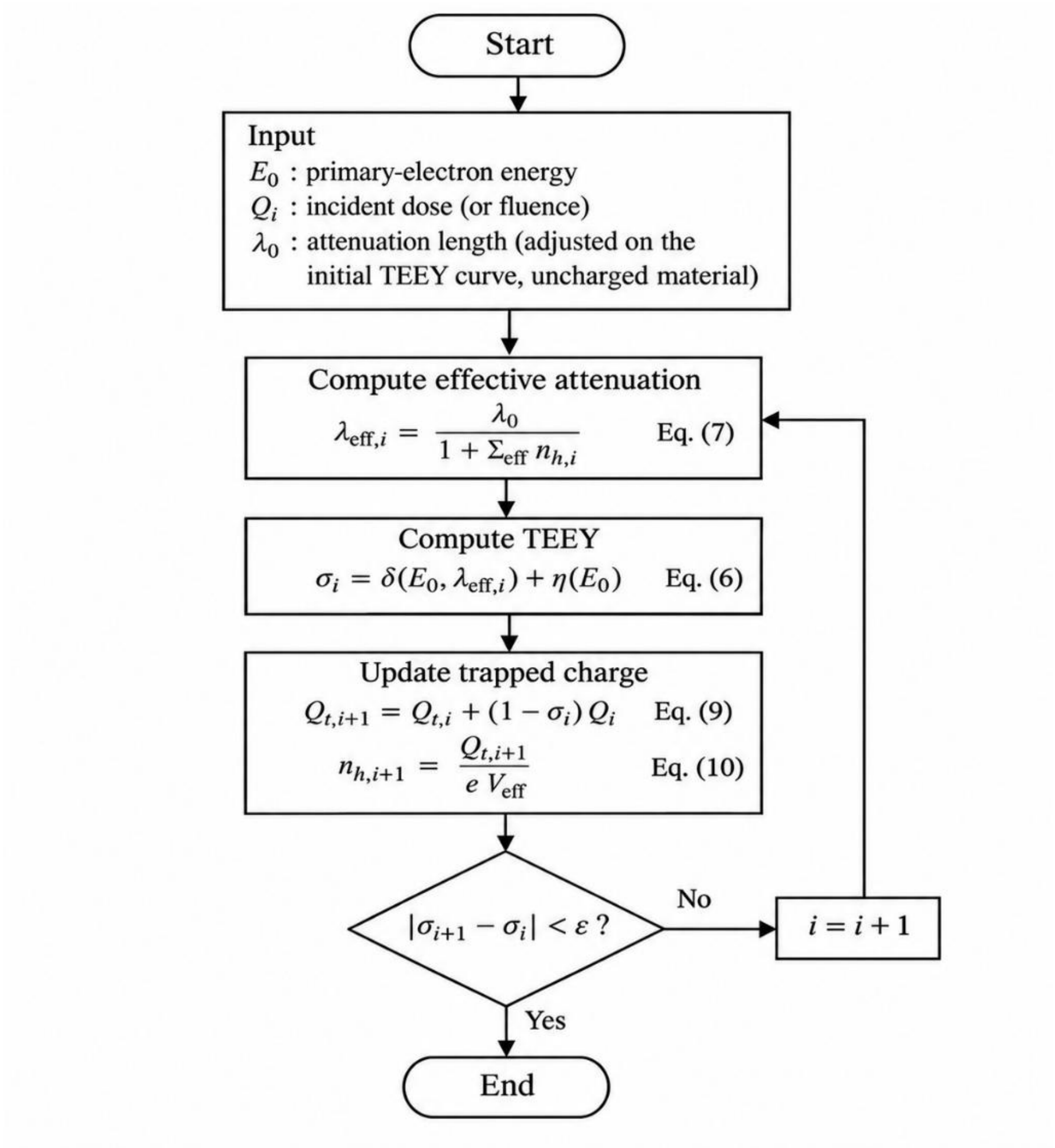

Figure 1. Flowchart of the charge-dependent TEEY calculation.

Charging may also modify electron trajectories in vacuum, the landing energy of incident electrons, and the surface escape barrier [4, 5]. The present model concerns secondary-electron transport inside the dielectric and escape at the interface. Therefore, the influence of electrostatic fields on trajectories in the vacuum region is not included explicitly, but it is generally implicitly considered on common particle-in-cell (PIC) or other simulation codes.

## 3. Application to polycrystalline $Al_2O_3$

The experimental TEEY curves of polycrystalline $Al_2O_3$ reported in Ref. [17] are reproduced in Fig. 2. The measurement procedure was designed to separate internal from external charging effects, so the observed TEEY decrease is attributed to internal charging. The yield was measured after seven successive irradiation pulses of 25 µs each, with approximately 2 pC injected per pulse, over an irradiated area of approximately S = 0.03 cm$^2$. The cumulative injected charges were 2, 4, 6, 8, 10, 12 and 14 pC. It should be noted that the measurements were performed on surfaces previously exposed to the atmosphere. Hydrocarbon and hydroxide contamination is therefore likely to be present on these surfaces.

The TEEY curve measured at $Q_{i,0}$ = 2 pC is taken as the reference state for the calculations.

For $Al_2O_3$, ρ = 3.95 g cm$^{-3}$, A/Z ≈ 101.96/50, and $E_g$ = 8.8 eV, corresponding to $\varepsilon_{pair}$ = 26.4 eV.

Secondary electrons are assumed to be generated isotropically; hence, at most half of the generated secondary electrons are initially directed towards the surface. Their transmission into vacuum also depends on the potential barrier at the solid–vacuum interface, as reflected in established descriptions of secondary-electron attenuation and escape [10, 11, 14]. In the effective treatment adopted here, the escape factor B is fixed at 0.5 when fitting the reference TEEY curve. The backscattered-primary contribution is fixed at b = η = 0.2, consistent with the representative value adopted in alumina charging models [16]. Its independence from incident energy is an approximation, particularly at low energy, where elastic reflection and the relative backscattered contribution can become important [4, 31].

With B = 0.5 and b = 0.2 fixed, only the reference attenuation length $\lambda_0$ is adjusted to fit the measured reference TEEY curve; the fitted value is $\lambda_0$ = 1.89 nm.

Figure 2 shows the measured TEEY curves and the corresponding reference Dionne calculation.

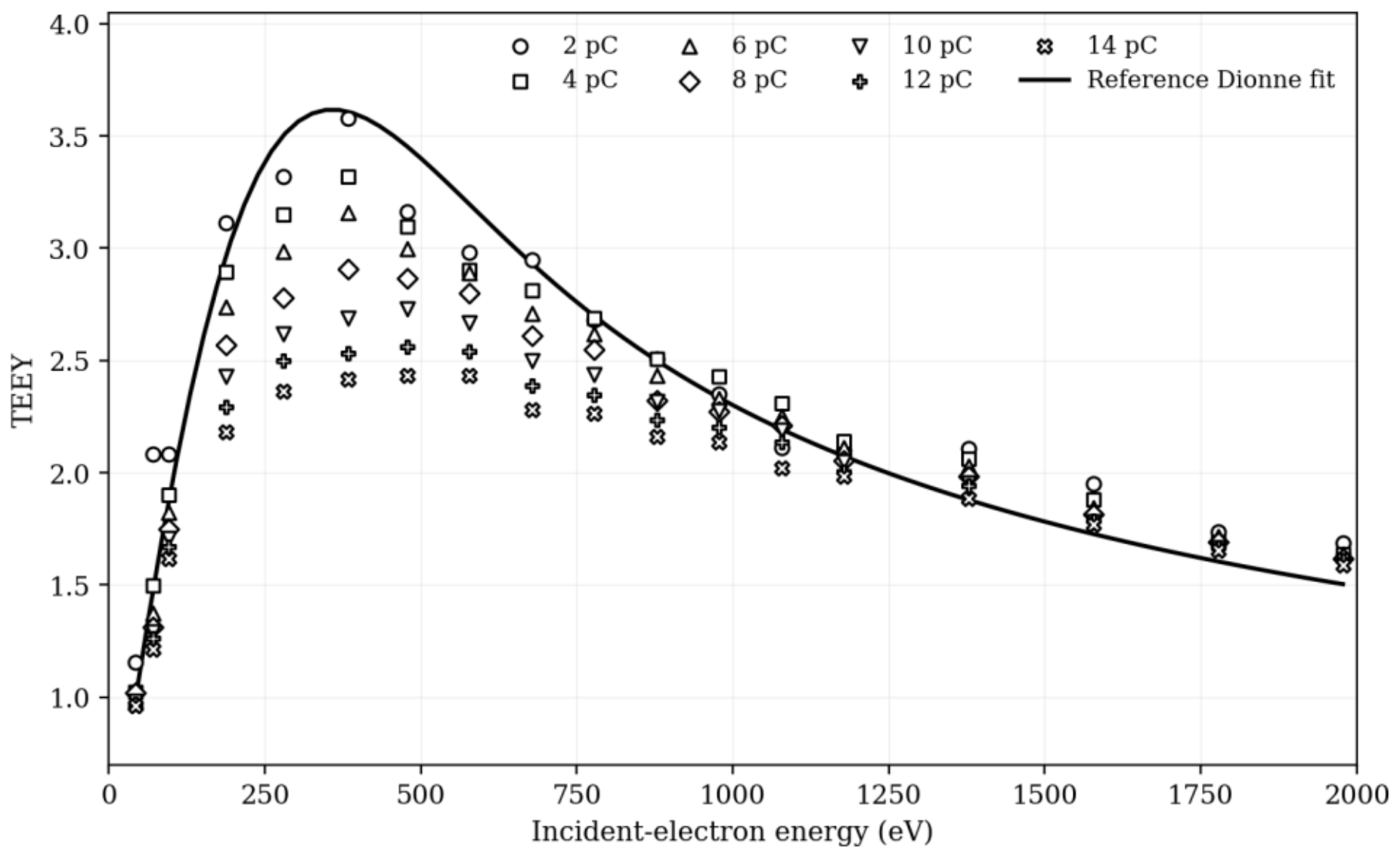

Figure 2. Complete experimental TEEY curves at all 18 measured energies (43–1978 eV) for the seven cumulative injected charges (2–14 pC). The black solid line represents the reference Dionne calculation (B = 0.5, b = 0.2, $\lambda_0$ = 1.89 nm).

Once the initial parameters, including $\lambda_0$, are fixed, the reference state is given by

$$TEEY(E_0, Q_i) = b + [\gamma_0 - b]\frac{\lambda_{eff}(E_0,Q_i)}{\lambda_0}\frac{1-exp[-R(E_0)/\lambda_{eff}(E_0,Q_i)]}{1-exp[-R(E_0)/\lambda_0]} \quad (11)$$

The effective cross section $\Sigma_{eff}$ describes the interaction of trapped positive charge with secondary electrons inside the solid. This cross section is determined from the measured variation of the TEEY with injected charge.

The change in trapped charge $Q_t$ with cumulative injected charge $Q_i$ follows from charge balance, with $D = Q_i/(eS)$, where S = 0.03 cm²:

$$\frac{dn_h}{dQ_i} = \begin{cases} \frac{\mathrm{TEEY}(E_0,Q_i)-1}{eS\lambda_0}, & \mathrm{TEEY} > 1, \\ 0, & \mathrm{TEEY} \leq 1. \end{cases} \quad (12)$$

For TEEY > 1, the measured decrease in yield with injected charge is represented by Eq. (11). Once $\lambda_0$ is fixed by the reference TEEY curve, $\Sigma_{eff}$ is the only parameter adjusted independently.

For this purpose, the calculation procedure outlined in the flowchart of Fig. 1 was employed in an iterative fitting procedure to determine the optimum effective cross section at selected primary-electron energies lying between the two crossover energies (TEEY = 1), for which the charge-induced variation in TEEY is appreciable. These energies, as indicated in Fig. 2, are 187, 280, 383, 478, 578, 678, 778, and 878 eV. The corresponding fitted curves are compared in Fig. 3.

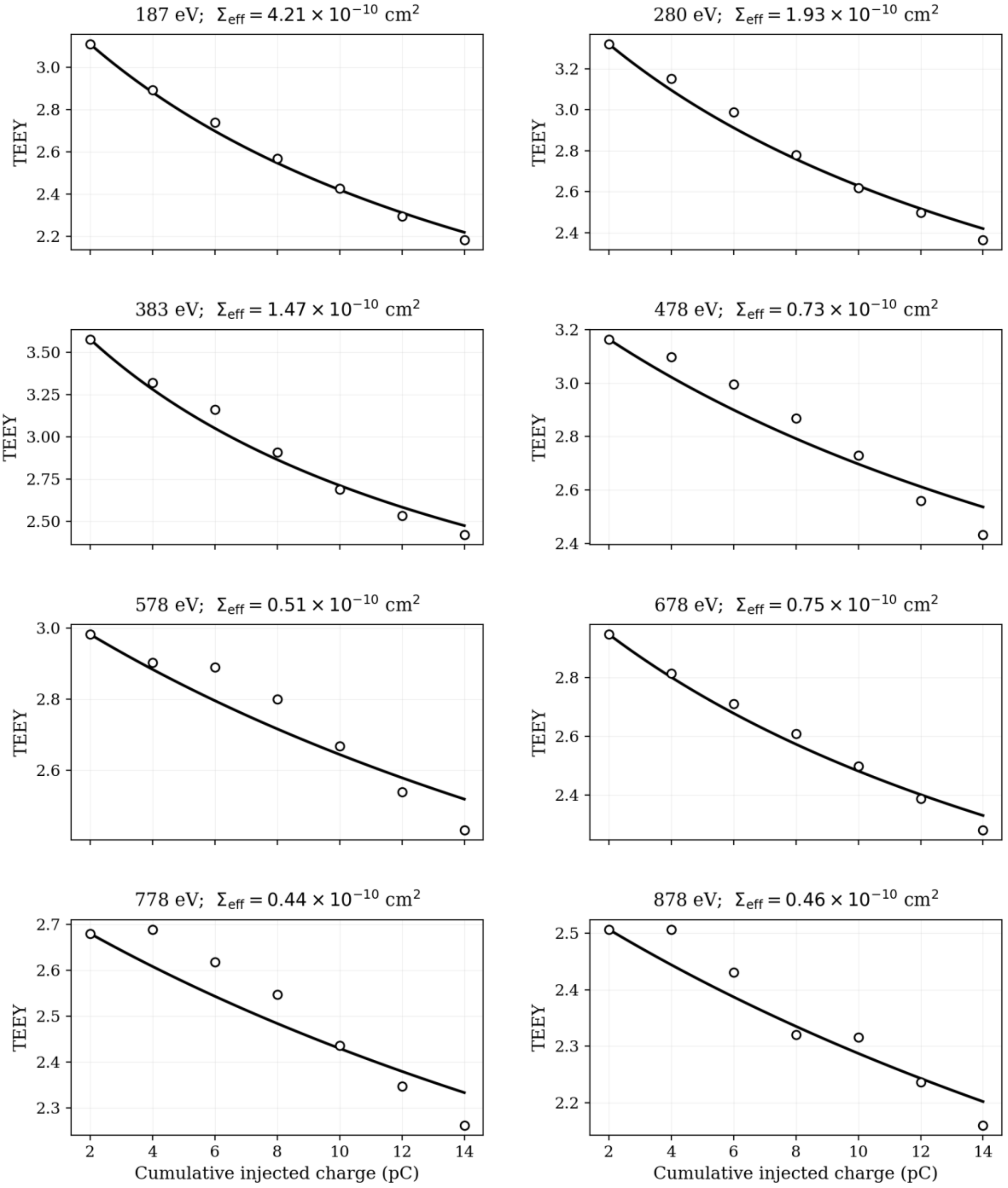

Figure 3. Measured charge-dependent TEEY (open symbols) and fitted model curves (solid lines) at 187, 280, 383, 478, 578, 678, 778 and 878 eV. Each panel reports the corresponding effective cross section.

The mean effective interaction cross section obtained from the eight energies is $\bar{\Sigma}_{eff} = 1.31 \times 10^{-10}$ cm², with a sample standard deviation $s_{\Sigma} = 1.29 \times 10^{-10}$ cm².

Because the energy distribution of secondary electrons is not expected to vary noticeably with incident energy $E_0$, a pronounced energy dependence of this effective parameter is not anticipated. The dispersion of the fitted values may arise from the assumptions of the reduced model and experimental uncertainties and in particular the variation of the electron beam diameter (and therefore S) as function of the incident electron energy [18]. The mean cross section is then used to predict the charge-dependent TEEY curves over the full incident-energy range.

Figure 4 compares the measured TEEY curves with nonlinear predictions obtained using $\bar{\Sigma}_{eff} = 1.31 \times 10^{-10}$ cm². All measured incident energies are included in this comparison. The quantitative agreement deteriorates as the injected charge increases, which may reflect both the approximations of the model and a decrease in experimental sensitivity during irradiation.

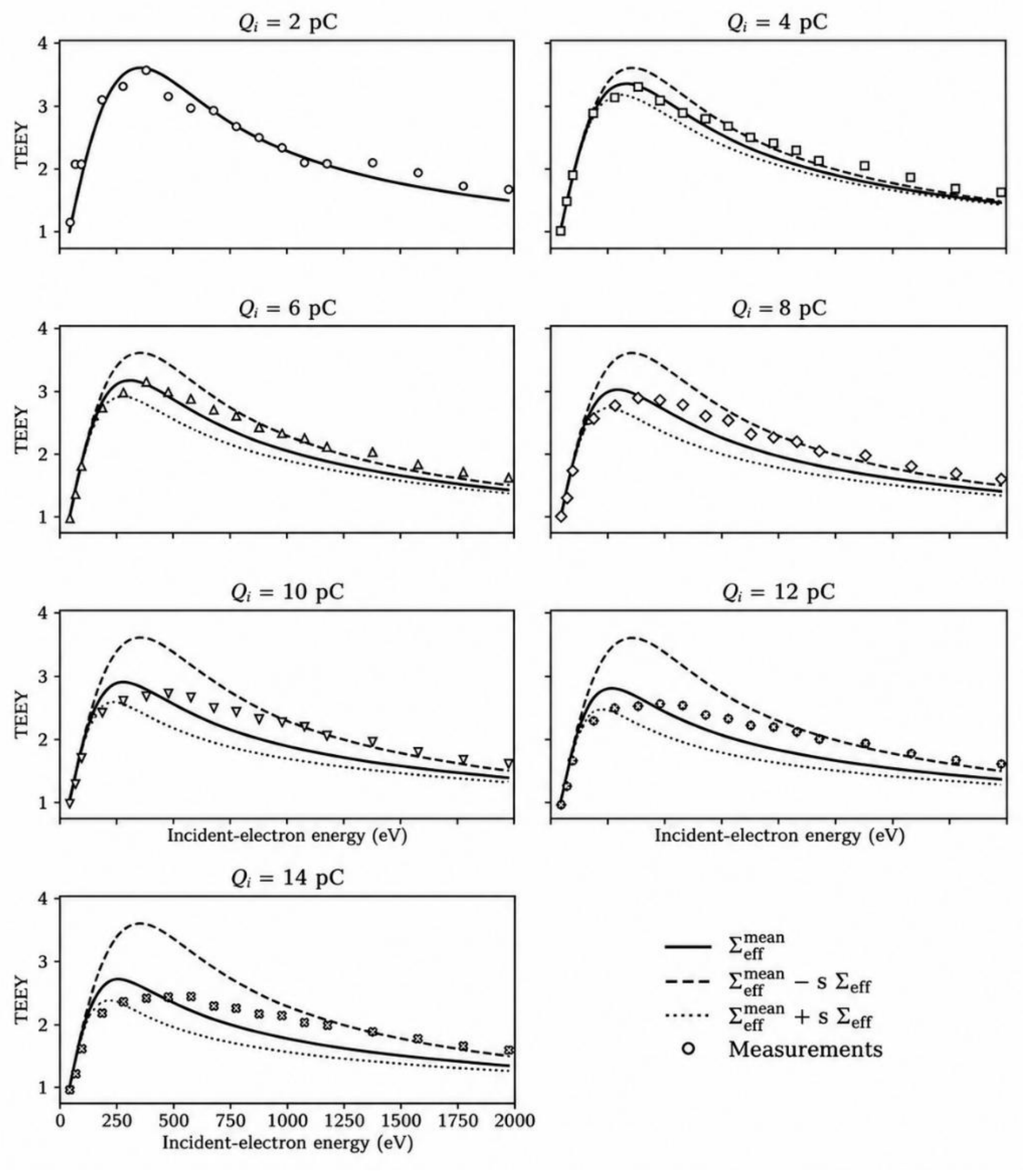

Figure 4. Comparison of the charge-dependent TEEY model with experimental measurements symbols at seven cumulative injected charges (2–14 pC) Solid line: $\bar{\Sigma}_{eff} = 1.31 \times 10^{-10}$ cm²; dashed line: $\bar{\Sigma}_{eff} - s_{\Sigma}$; dotted line: $\bar{\Sigma}_{eff} + s_{\Sigma}$.

The nonlinear model is then used beyond the measured range of injected charge, without further adjustment of its parameters. Figure 5 shows the calculated TEEY curves and their variation with charge at four representative primary energies.

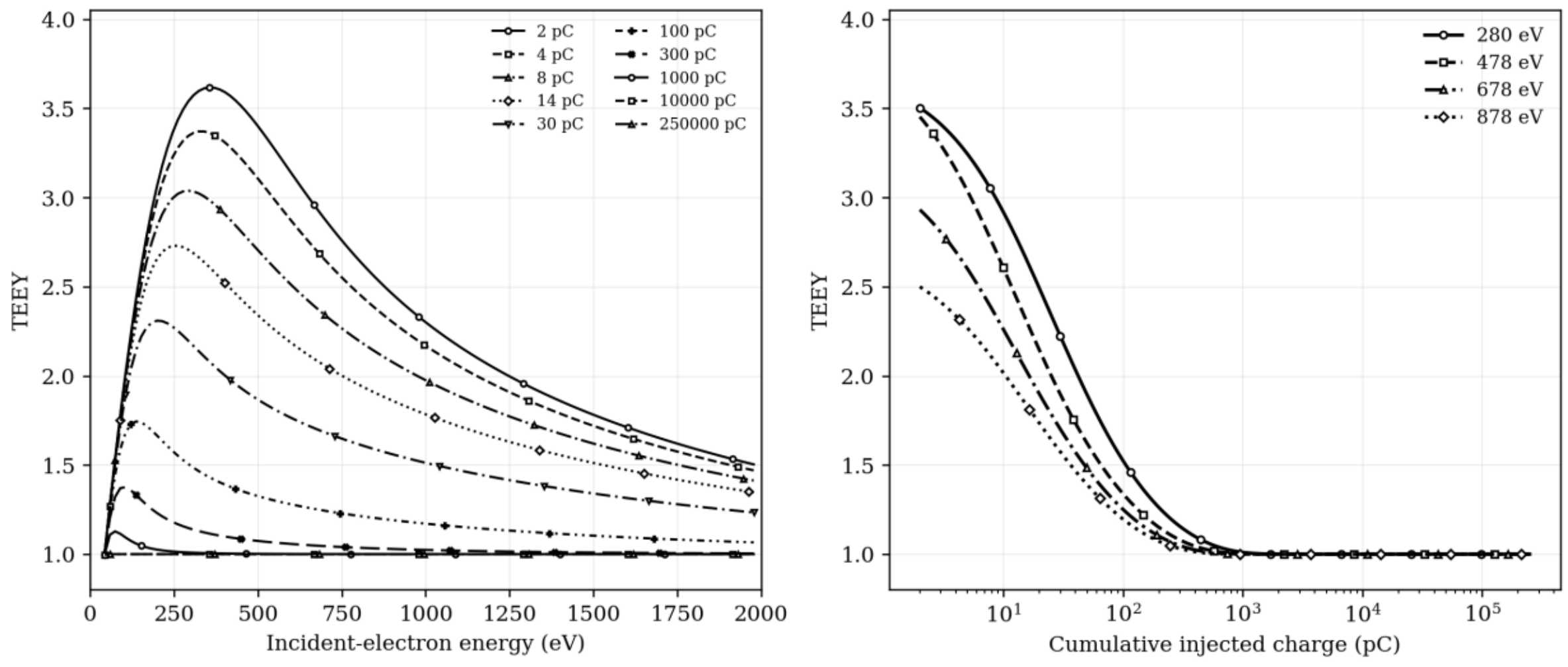

Figure 5. Full nonlinear TEEY predictions approaching practical saturation. Left: simulated TEEY curves with increasing injected charge, beginning at 2 pC. Right: TEEY versus cumulative injected charge $Q_i$ at 280, 478, 678 and 878 eV.

The corresponding trapped-hole density and effective attenuation length, calculated using the mean cross section, are shown in Fig. 6 for the same four energies. Together, Figs. 5 and 6 illustrate the internal feedback mechanism. As long as TEEY exceeds unity, the material traps positive charge and the trapped-hole density increases at a rate proportional to TEEY − 1. The associated decrease in effective attenuation length progressively reduces emission. Consequently, the growth rate of trapped charge tends to zero as the system approaches the limiting condition TEEY = 1.

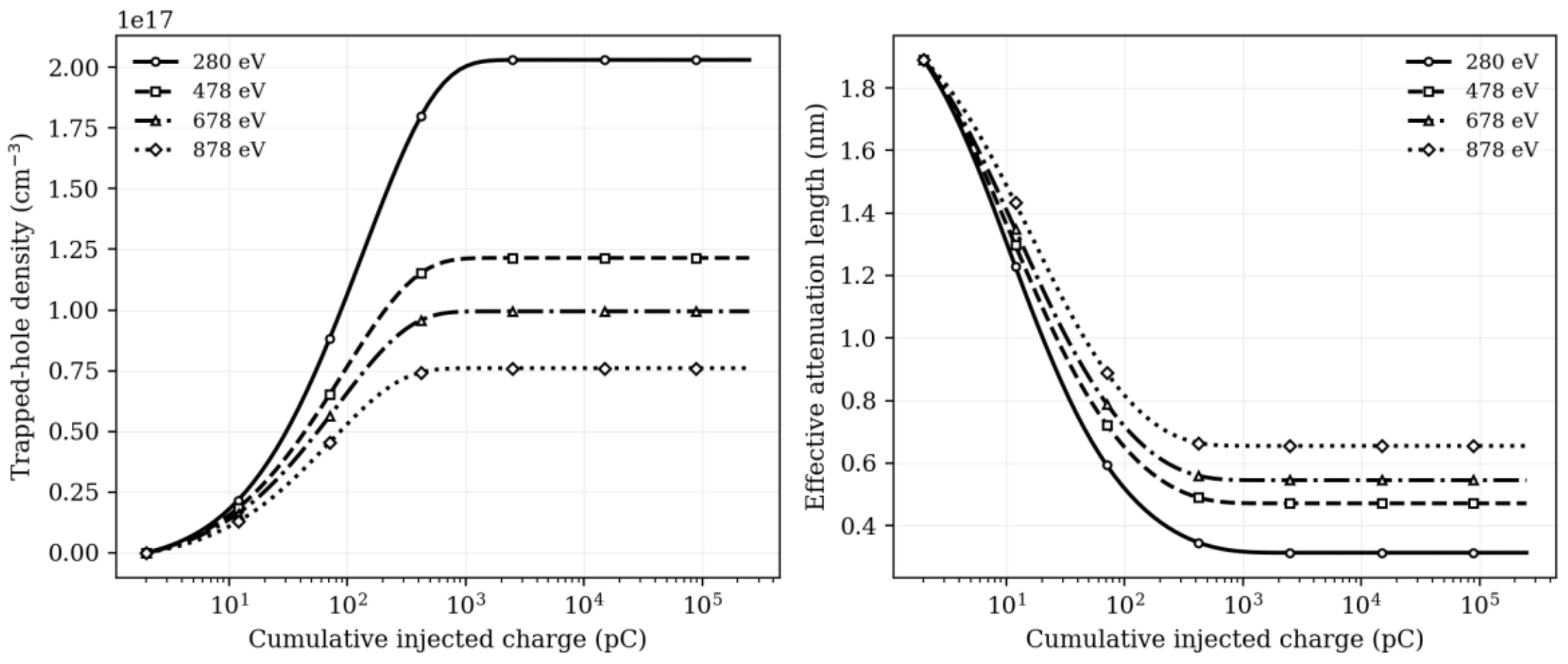

Figure 6. Trapped-hole density $n_h$ and effective attenuation length $\lambda_{eff}$ versus cumulative injected charge $Q_i$ at 280, 478, 678 and 878 eV. All curves use $\bar{\Sigma}_{eff} = 1.31 \times 10^{-10}$ cm².

## 4. Analytical approximation of charge-dependent TEEY

The above results describe the influence of internal charging on the TEEY of alumina. However, repeated numerical integration of the coupled charge-balance equation may be inconvenient in more comprehensive and complex electron-irradiation simulations. An analytical approximation of the charge-dependent emission law is therefore required.

At fixed incident-electron energy $E_0$, the limiting condition TEEY = 1 defines the attenuation length $\lambda_{sat}$. Substitution into the Dionne relation yields a transcendental equation because $\lambda_{sat}$ occurs both as a multiplicative factor and in the exponential; it cannot be solved by elementary algebra.

Introducing $u = R(E_0)/\lambda_{sat}$, the saturation equation becomes $u = c[1 - \exp(-u)]$, where $c = BE_0/[\varepsilon_{pair}(1 - b)]$. This equation can be rearranged as $w \exp(w) = z$ and solved analytically using the Lambert W function, defined by $W(ze^z) = z$..

The nonzero saturation solution is obtained from the principal real branch $W_0$, giving $\lambda_{sat}$ explicitly in terms of incident energy and the Dionne-law parameters:

$$\frac{1}{\lambda_{sat}(E_0)} = \frac{1}{R(E_0)}\{\frac{BE_0}{\varepsilon_{pair}(1-b)} + W_0[\frac{-BE_0}{\varepsilon_{pair}(1-b)} exp(\frac{-BE_0}{\varepsilon_{pair}(1-b)})]\} \tag{13}$$

The intermediate algebra and branch selection are detailed in the Appendix A. Once $\lambda_{sat}$ is determined, the evolution of $\lambda_{eff}$ from its reference value toward saturation is represented using the initial excess yield:

$$\Delta TEEY_0(E_0) = b + \frac{BE_0}{\varepsilon_{pair}}\frac{\lambda_0}{R(E_0)}[1 - exp(\frac{-R(E_0)}{\lambda_0})] - 1 \tag{14}$$

The characteristic length $\ell_{ch}(E_0)$ controls the curvature of the analytical charging relation:

$$l_{ch}(E_0) = \frac{-1}{\lambda_{sat}^{-1}(E_0) - \lambda_0^{-1}} + \frac{BE_0\lambda_0^2}{\varepsilon_{pair}\Delta TEEY_0(E_0)R(E_0)}[1 - (1 + \frac{R(E_0)}{\lambda_0})exp(\frac{-R(E_0)}{\lambda_0})] \tag{15}$$

The length $\ell_{ch}$ determines the curvature of the analytical relation while preserving the reference yield and the saturation limit.

The corresponding characteristic injected charge $Q_{ch}(E_0)$, expressed in coulombs, is

$$Q_{ch}(E_0) = \frac{eS\lambda_0}{\Sigma_{eff}\Delta TEEY_0(E_0)}[\lambda_{sat}^{-1}(E_0) - \lambda_0^{-1}] \tag{16}$$

For Eq. (17), $\Delta\kappa(E_0) = \lambda_{sat}^{-1}(E_0) - \lambda_0^{-1}$.

$$\frac{1}{\lambda_{eff}(E_0,Q_i)} = \frac{1}{\lambda_{sat}(E_0)} + \frac{1+l_{ch}\Delta\kappa}{l_{ch}} W_0[\frac{-l_{ch}\Delta\kappa}{1+l_{ch}\Delta\kappa} exp(\frac{-(Q_i - Q_{i,0})/Q_{ch} + l_{ch}\Delta\kappa}{1+l_{ch}\Delta\kappa})]$$

Where $\Delta\kappa(E_0) = \lambda_{sat}^{-1}(E_0) - \lambda_0^{-1}$. (17)

Substituting the effective attenuation length into the Dionne emission law gives the final analytical expression for TEEY:

$$TEEY_{an}(E_0, Q_i) = b + \frac{BE_0}{\varepsilon_{pair}} \frac{\lambda_{eff}(E_0,Q_i)}{R(E_0)} [1 - exp(\frac{-R(E_0)}{\lambda_{eff}(E_0,Q_i)})] \quad (18)$$

The TEEY is thus given directly as a function of incident energy $E_0$ and injected charge $Q_i$.

Figure 7 compares the analytical approximation with the nonlinear calculation at larger injected charges, up to the limiting condition TEEY = 1. The same mean cross section and four representative incident energies are retained. The close agreement persists during extrapolation towards saturation.

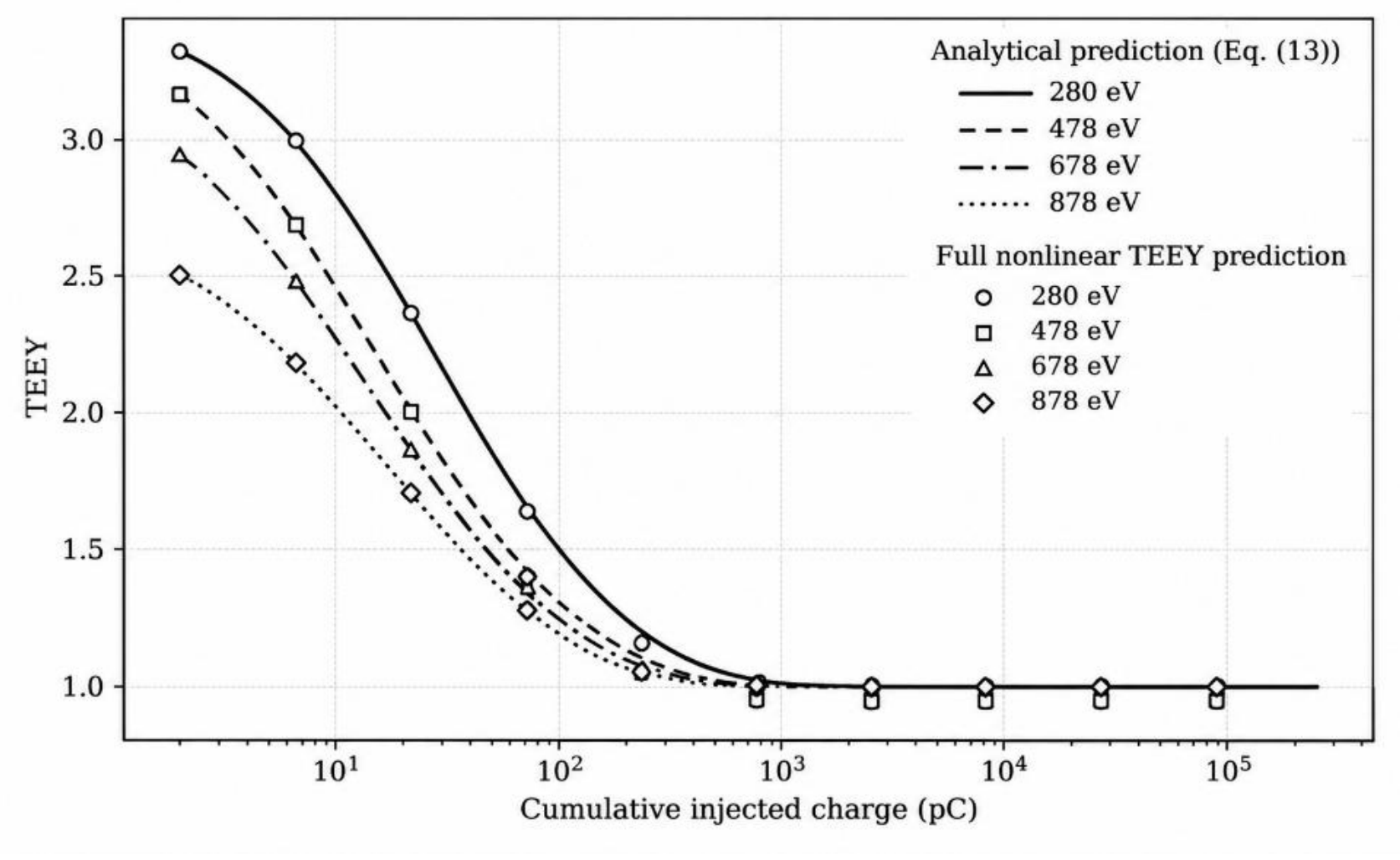

Figure 7. Closed-form analytical predictions (open symbols) compared with full nonlinear results (lines) up to practical saturation at 280, 478, 678 and 878 eV, using $\bar{\Sigma}_{eff} = 1.31 \times 10^{-10}$ cm².

## 5. Discussion

Both the nonlinear model and its analytical approximation describe the measured dependence of TEEY on injected charge or, for a specified incident current, on irradiation time, provided that the reference yield and the effective charge–electron interaction cross section are known. The measured reference curve is adopted as the initial condition, assuming that its modification by charging is sufficiently small. However, obtaining a truly charge-free TEEY curve for an insulator is difficult. Dennison et al. [6] and Hoffmann et al. [34] reported irradiation-induced changes even for charge fluences in the fC range. Furthermore, conventional measurements do not generally provide an independent determination of injected or trapped charge. By contrast, Kelvin-probe measurements, such as those of Mata et al. [35] and Cavalca de Almeida et al. [36], allow the surface potential or charge to be monitored together with the electron-emission response.

The effective interaction cross section is a central parameter in the present formulation. Indeed, once $\lambda_0$ has been fixed by fitting to experimental measurements or other available data, the effective cross section remains the sole adjustable parameter. This constitutes both the principal advantage of the model and its inherent limitation. Reported values for charge-dependent electron emission and carrier recombination span several orders of magnitude. For example, authors in reference [9] inferred values of a few $10^{-11}$ cm² for polycrystalline MgO; Boughariou et al. [20]

reported values of order $10^{-10}$ cm² for positively charged spinel; and Said et al. [19] obtained values of order $10^{-9}$ cm² for sapphire and polycrystalline alumina. Although these effective cross sections were determined for different materials using different approaches, they indicate charge-electrons interactions of broadly comparable magnitude. Within the present interpretation, the Coulomb attraction due to trapped positive charge affects the transport of low-energy secondary electrons through local electric fields and electron–hole recombination. Such interactions may therefore produce additional attenuation beyond the elastic and inelastic processes represented by the reference attenuation length.

The magnitude of this additional attenuation is illustrated in Fig. 6. At 280 eV, the effective attenuation length decreases by approximately a factor of five as saturation is approached. The initial TEEY curve depends on the physicochemical and structural properties of the dielectric. During continued irradiation, however, the evolution increasingly depends on the ability of the material to retain positive charge and on the associated high effective interaction cross section. Thus, the reference TEEY determines the initial emission response, whereas charge trapping controls its subsequent evolution. Determining the effective cross section experimentally requires a measurement procedure that separates internal charging from external electrostatic effects. Such measurements are demanding and are not necessarily available for all materials used in multipactor or electrostatic-discharge simulations.

To assess this uncertainty, calculations at 280 eV were performed for $\Sigma_{eff} = 10^{-11}$, $10^{-10}$, and $10^{-9}$ cm² (Fig. 8), with identical reference TEEY parameters. Only the effective interaction cross section was varied. Figure 8 shows the resulting evolution of TEEY with injected charge.

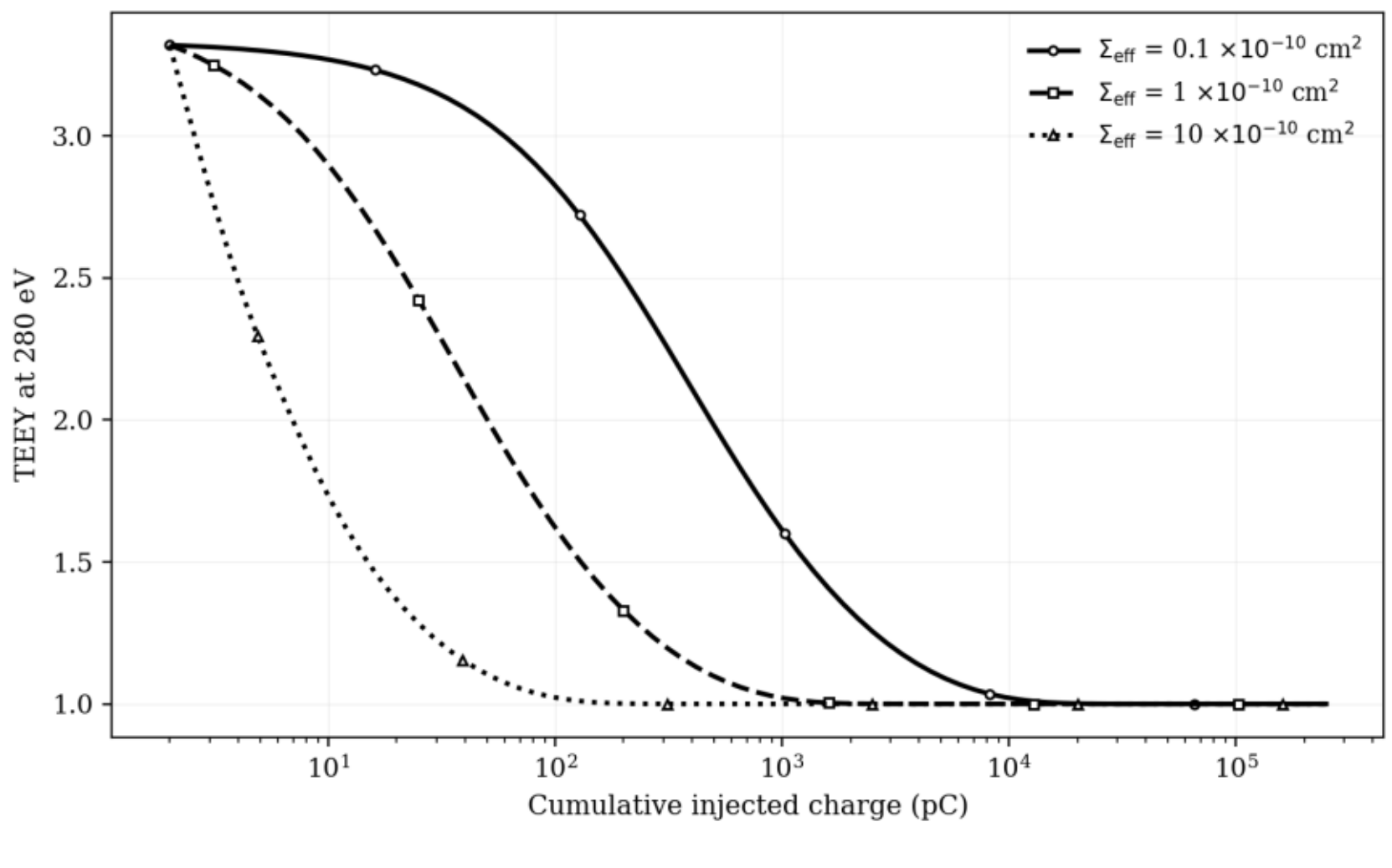

Figure 8. Effect of effective charge–electron interaction cross section on the TEEY curve at 280 eV. TEEY versus cumulative injected charge $Q_i$ for $\Sigma_{eff}$ = 0.1, 1 and 10 × $10^{-10}$ cm².

All three calculations follow the same charging mechanism, but on different injected-charge scales. At fixed incident energy and fixed reference yield, the charge-balance equation depends on injected charge through a scaled variable proportional to $\Sigma_{eff}$ $Q_i$. Although the time scale of saturation can be essential in some applications, uncertainty in $\Sigma_{eff}$ may be less critical when the purpose is a semi-quantitative or qualitative interpretation. Despite its simplifying assumptions, the present charge-dependent model therefore offers an improvement over a static emission law that cannot evolve with accumulated charge. The magnitude of the observed variations in TEEY suggests that the underlying physical mechanism should not be neglected in electron-irradiation simulations of dielectric materials. The proposed model provides a relatively simple means of accounting for this effect and can readily be incorporated into existing simulation codes. Although the determination of the effective interaction cross section remains subject to uncertainty, the model may nevertheless provide useful insight into the physical processes involved in a variety of electron-irradiation problems, particularly those associated with multipactor discharges.

## 6. Conclusion

The Dionne formulation has been extended to account for the influence of trapped positive charge on the total electron emission yield of insulating materials. By introducing an additional attenuation of secondary electrons during their transport towards the surface, a self-consistent coupling between electron emission and charge accumulation is established.

An analytical approximation of the nonlinear model has also been derived. It provides good agreement with the numerical solution while considerably reducing the computational effort required for its implementation.

The proposed approach may be particularly useful in simulations of multipactor discharges involving dielectric materials. Although such simulations can account for the electrostatic field associated with surface charging, they generally neglect its influence on secondary-electron transport within the dielectric. The present formulation offers a relatively simple means of including this effect, without resorting to a detailed charge-transport calculation. Despite the uncertainties associated with the determination of the effective interaction cross section, it may thus provide useful phenomenological insight into electron-irradiation processes in dielectric materials.

## Acknowledgments

This research was supported by the ONERA–CNES joint research project "PIC MULTIPACTOR."

## Appendix A

## Analytical derivation

### General analytical derivation

Let $Y(E_0, \lambda)$ denote the reference TEEY curve, whether measured or calculated. According to the Dionne model [10, 11], at fixed incident energy $E_0$, the emission yield is

Here $R = R(E_0)$, $B$ is the escape coefficient, $b$ the backscattered contribution, $\varepsilon_{\text{pair}}$ the pair-creation energy, and $\lambda_0$ the reference attenuation length.

$$Y(\lambda; E_0) = b + \frac{BE_0}{\varepsilon_{\text{pair}}} \frac{\lambda}{R}(1 - e^{-R/\lambda}) \quad \text{(A1)}$$

At the reference state, $Y_0 = Y(E_0, \lambda_0)$, corresponding to the uncharged dielectric.

The effective cross section $\Sigma_{\text{eff}}$ and trapped-hole density $n_h$ are related to the inverse attenuation length by

$$\kappa \equiv \frac{1}{\lambda_{\text{eff}}} = \frac{1}{\lambda_0} + \Sigma_{\text{eff}} n_h, \qquad n_h = \frac{Q_t}{eS\lambda_0} \quad \text{(A2)}$$

With $Q_t$ the trapped charge, $Q_i$ the cumulative injected charge, $S$ the irradiated area, and $e$ the elementary charge, the positive-charge balance gives

$$\frac{dQ_t}{dQ_i} = Y - 1, \qquad \frac{dn_h}{dQ_i} = \frac{Y-1}{eS\lambda_0} \quad \text{(A3)}$$

Set $\kappa_0 = 1/\lambda_0$, $x = \kappa - \kappa_0 = \Sigma_{\text{eff}} n_h$, and $g(x) = Y[1/(\kappa_0 + x)] - 1$. Taking $x(0) = 0$ and $Q_i(0) = 0$ gives

$$\frac{dx}{dQ_i} = \frac{\Sigma_{\text{eff}}}{eS\lambda_0} g(x), \qquad x(Q_{i,0}) = 0 \quad \text{(A4)}$$

This relation applies to the positive-charging case, $Y > 1$.

At saturation, $Y = 1$. Introducing $A = BE_0/\varepsilon_{\text{pair}}$, $c = A/(1 - b)$, and $u = R/\lambda_{\text{sat}}$ gives

$$1 - b = A\frac{1-e^{-u}}{u}, \qquad u = c(1 - e^{-u}) \quad \text{(A5)}$$

Rearranging yields $(u - c)\exp(u - c) = -c\exp(-c)$.

This is the form of the Lambert $W$ function, $W(ze^z) = z$. The finite saturation solution is

$$\kappa_{\text{sat}} = \frac{c+W_0(-ce^{-c})}{R}, \qquad \lambda_{\text{sat}} = \frac{R}{c+W_0(-ce^{-c})} \quad \text{(A6)}$$

For $c > 1$, the principal branch $W_0$ gives $u > 0$; the other real branch returns the trivial root $u = 0$. The positive-charging trajectory requires $Y(\lambda_0) > 1$ and $\lambda_{\text{sat}} < \lambda_0$.

Define $\Delta Y_0 = Y(\lambda_0) - 1$ and $\Delta\kappa = \kappa_{\text{sat}} - \kappa_0$. Since $g(0) = \Delta Y_0$ and $g(\Delta\kappa) = 0$, a rational approximation preserving both endpoints is

$$g(x) \simeq \tilde{g}(x) = \Delta Y_0 \, \frac{1 - x/\Delta\kappa}{1 + \ell_{\text{ch}} x} \quad \text{(A7)}$$

Differentiating the Dionne expression with respect to $\kappa$ at $\kappa_0$ gives

$$g'(0) = -\frac{B E_0 \lambda_0^2}{\varepsilon_{\text{pair}} R}\left[1 - \left(1 + \frac{R}{\lambda_0}\right) e^{-R/\lambda_0}\right] \quad \text{(A8)}$$

Matching the derivative of the rational approximation at $x = 0$ gives

$$\ell_{\text{ch}} = -\frac{1}{\Delta\kappa} - \frac{g'(0)}{\Delta Y_0} \quad \text{(A9)}$$

The approximation is applied for $0 \le x \le \Delta\kappa$, with $1 + \ell_{\text{ch}} x > 0$.

Introduce the characteristic injected charge $Q_{\text{ch}}$ and dimensionless variables

$$Q_{\text{ch}} = \frac{e S \lambda_0 \Delta\kappa}{\Sigma_{\text{eff}} \Delta Y_0}, \quad y = \frac{x}{\Delta\kappa}, \quad a = \ell_{\text{ch}} \Delta\kappa, \quad q = \frac{Q_i}{Q_{\text{ch}}} \quad \text{(A10)}$$

Substitution into the charge-balance equation yields a separable differential equation. Integrating with $y(0) = 0$ gives

$$\frac{dy}{dq} = \frac{1 - y}{1 + a y}, \qquad q = -(1 + a)\ln(1 - y) - a y \quad \text{(A11)}$$

Setting $z = 1 - y$ and rearranging into the form $w\exp(w) = z$ gives, using the principal branch $W_0$,

$$\kappa = \kappa_{\text{sat}} + \frac{1 + a}{\ell_{\text{ch}}} W_0\left[-\frac{a}{1 + a} e^{-(q + a)/(1 + a)}\right] \quad \text{(A12)}$$

When $\ell_{\text{ch}} = 0$, the continuous limit is $\kappa = \kappa_{\text{sat}} - \Delta\kappa \exp(-q)$. Substitution into the original emission law gives

$$Y_{\text{an}}(E_0, Q_i) = b + \frac{B E_0}{\varepsilon_{\text{pair}}} \frac{1 - e^{-R\kappa}}{R\kappa} \quad \text{(A13)}$$

## Example of numerical implementation for $Al_2O_3$

For $E_0 = 280$ eV, the reference charge is $Q_{i,0} = 2$ pC and the target cumulative injected charge is $Q_i = 20$ pC, corresponding to an additional injected charge of 18 pC.

| | |
|---|---|
| $E_0$ | 280 eV |
| $B$ and $b$ | 0.5; 0.2 |
| $\varepsilon_{\mathrm{pair}}$ | 26.4 eV |
| $\lambda_0$ | 1.89 nm |
| $\rho$ | 3.95 g cm$^{-3}$ |
| $A/Z$ | 101.96/50 |
| $\Sigma_{\mathrm{eff}}$ | $1.31 \times 10^{-10}$ cm$^2$ |
| $S$ | 0.03 cm$^2$ |
| Qi(t$_0$) and Q(t$_0$+t) | 2 pC; 20 pC |

Using the Castaing range expression with $E_0$ in keV gives

$$R(\mu\mathrm{m}) = 3.30 \times 10^{-2}\,\frac{A/Z}{\rho}\,[E_0(\mathrm{keV})]^{1.7} \quad \text{(A14)}$$

With $E_0 = 0.28$ keV, $R = (3.30 \times 10^{-2})(101.96/50)(0.28)^{1.7}/3.95$ µm $= 1.96$ nm.

The parameters give $A = BE_0/\varepsilon_{\mathrm{pair}} = 5.3$, $c = A/(1-b) = 6.63$, and $\kappa_0 = 1/\lambda_0 = 0.53$ nm$^{-1}$.

Hence $Y_0 = 3.5$ and $\Delta Y_0 = 2.5$.

The Lambert argument is $-c\exp(-c) = -8.76 \times 10^{-3}$.

From Eq. (A6), $\kappa_{\mathrm{sat}} = 3.38$ nm$^{-1}$, $\lambda_{\mathrm{sat}} = 0.3$ nm, and $\Delta\kappa = 2.85$ nm$^{-1}$.

Equations (A8) and (A9) give $g'(0) = -2.68$ nm and $\ell_{\mathrm{ch}} = 0.72$ nm. Therefore, $a = \ell_{\mathrm{ch}}\Delta\kappa = 2.06$.

Converting lengths and cross-section units consistently, Eq. (A10) gives $Q_{\mathrm{ch}} = 79.06$ pC.

For $q = (20-2)/Q_{\mathrm{ch}} = 0.23$, Eq. (A12) gives $\kappa = 1.03$ nm$^{-1}$ and $\lambda_{\mathrm{eff}} = 0.97$ nm.

The final emission yield is:

$$\mathrm{TEEY}_{\mathrm{an}}(280\,\mathrm{eV}, 20\,\mathrm{pC}) \simeq 2.48 \quad \text{(A15)}$$